# Seismic precursory pattern before a cliff collapse and critical-point phenomena

D. Amitrano*, J.R Grasso[#], G. Senfaute*

\* LAEGO-INPL-INERIS, Nancy, France
[#] LGIT, Grenoble, France and USGS Menlo Park, USA

**Abstract:**

We analyse the statistical pattern of seismicity before a 1-2 $10^3$ m$^3$ chalk cliff collapse on the Normandie ocean shore, Western France. We show that a power law acceleration of seismicity rate and energy in both 40 Hz-1.5 kHz and 2 Hz-10kHz frequency range, is defined on 3 order of magnitude, within 2 hours from the collapse time. Simultaneously, the average size of the seismic events increases toward the time to failure. These in-situ results are derived from the only station located within one rupture length distance from the rock fall rupture plane. They mimic the "critical point" like behavior recovered from physical and numerical experiments before brittle failures and tertiary creep failures. Our analysis of this first seismic monitoring data of a cliff collapse suggests that the thermodynamic phase transition models for failure may apply for cliff collapse.

**Introduction**

Recent advances in monitoring techniques, including GPS and synthetic aperture radar for surface displacements and seismic sensors for slide-quakes allow to investigate the time dependence in slope failure processes. Standard stability analyses, which insert geomechanics field data in numerical models of failure to predict the distance to failure for a given slope are, for most of them, time independent estimates which by nature do not account for acceleration in slope movements [*Stead et al.*, 2001]. While it is obvious that monitoring data are critical for being able to successfully forecast landslide occurrence, they are also critical for understanding how natural slopes collapse.

Most of the case studies that report accelerating patterns prior to slope movements use displacement or deformation rates [*Bhandari*, 1988; *Zvelebil and Moser*, 2001; *Petley et al.*, 2002]. For some of them the acceleration is either an exponential or power law toward time to collapse. Petley et al. [2002] suggested the exponential law is observed for landslides in ductile materials. The power law accelerating displacement of the slope prior to the collapse is proposed to be analog to the final stage of the tertiary creep as observed on lab scale experiments [*Saito and Uezawa*, 1961; *Kennedy and Niermeyer*, 1971; *Voight*, 1988]. Voight [1988] suggest this power law acceleration can be recovered when using other variables including, strain, seismicity rate, or seismic energy released and generalizes damage mechanics laws but it has never been reported before any landslide types up to now. More recently, [*Helmstetter et al.*, 2004] provide a physical basis for the phenomenological power law for landsliding displacement acceleration based on a slider block model using a rate- and velocity- friction law established in the laboratory [*Scholz*, 1998]. An alternative modelling strategy consists of viewing this power law accelerating micro-damage event as forerunners of the macroscopic brittle failure which is suggested to be analog to a thermodynamics phase transition [*Buchel and Sethna*, 1997; *Zapperi et al.*, 1997; *Sornette and Andersen*, 1998; *Kun and Herrmann*, 1999]. This acceleration of brittle damage before failure is sometimes reported during lab scale experiments that use acoustic emission measurements during the fracturing of brittle heterogeneous material [*Guarino et al.*, 1998; *Johansen and Sornette*, 2000; *Nechad et al.*, 2005]. Nonetheless many other experiments do not reproduce the patterns predicted by statistical physics model before brittle failure and the applicability of these brittle failure models to the earth crust fracturing is still debated, e.g. the so-called critical point hypothesis for earthquakes [*Bufe and Varnes*, 1993; *Jaume and Sykes*, 1999; *Zoller and Hainzl*, 2002].

The acoustic emission (AE) tool has been extensively used at laboratory rock sample scale [for a review see *Lockner*, 1993] and, at an intermediate scale between the lab scale and the large tectonic earthquakes, for studies of seismicity and rockburst in mines or tunnels [e.g.*Obert* 1977; *Nicholson*, 1992]. A few applications of AE monitoring for slope stability are related to either open mine, quarry, landslides or volcano flanks [e.g. *McCauley*, 1976; *Hardy and Kimble*, 1991].

In this work, we test which precursory seismic pattern can be recovered before the failure of a natural cliff.

**Data**

The studied cliff is a natural chalk cliff, Mesnil-Val, Haute Normandie, France, where rock falls driven by sea erosion recurrently occur. The average coastline recession rate in this region is of the order of 0.5-1 m/yr. The choice of the monitoring site location was driven by empirical expert advices based on geological and structural configurations, which were considered to be prone for a cliff collapse to occur [*Senfaute et al.*, 2003]. The monitored sub-vertical cliff face is a NE-SW face, 50 m height and 60 m length. The cliff is made of two main horizontally layered chalk series, from Cenomanien and Turonian age respectively [*Mortimore*, 2001; *Senfaute et al.*, 2003]. Physical and mechanical properties of the chalk rock, as derived from laboratory test, are in the range 0.42-0.45 for porosity, 1.51- 1.71 for dry density, 4.4-8.4 GPa for Young modulus, 3.5-5.3 MPa for uniaxial strength, 0.15-0.25 for Poisson ratio, 2050-2600 m/s and 1000-1160 m/s for P and S wave velocity, respectively [*Senfaute et al.*, 2005].

A network of 5 seismic stations was installed on this site (figure 1a). Due to the strong attenuation of the signal in the high porosity chalky rocks, a maximum 50 meters sensors spacing was chosen, corresponding to 90% in amplitude decrease from calibration tests [*Senfaute et al.*, 2003]. Five stations were cemented, two within 10 m deep vertical boreholes, drilled form the top of the cliff at 10 m from the cliff edge, and three within horizontal boreholes, drilled perpendicular to the cliff face at a 6 m depth. Each seismic station is constituted by a geophone (40 Hz-1.5 kHz) and an accelerometer (2 Hz – 10 kHz) both connected to a 40/60 dB preamplifier and to a band-pass filter (170 Hz-10 kHz). All the sensors were connected to a digital acquisition system (40 kHz, 16 bits) which is continuously scanning the seismic signal on all the channels. When a given trigger threshold is reached on any channel, the signals from all sensors are recorded simultaneously for a duration of 0.35 s with a 0.05 s pre-trigger time. The complete network starts operating in January 2002.

A significant change in seismic recording style and a rate increase are observed during high tides, due to the effect of wave on the face of the cliff [*Senfaute et al.*, 2003]. It prevents us from comparing the seismicity during high and low tide period. During the low tide period, the mean seismicity rate over January-June 2002 was 1.7 event/day. A cliff collapse occurred on $23^{rd}$ June 2002 at the center of the monitored zone (figure 1a). The estimated rock fall volume was 1-2 $10^3$ $m^3$. Above normal seismicity rate was observed before the cliff collapse only on the A4 station where 200 events were recorded during 2 hours [*Senfaute et al.*, 2003]. A4 station is located at nearly 5 m from the rockfall rupture plane and is the only one which recorded measurable seismicity (figure 1b). The mechanisms involved in the collapse have been proposed to be both shearing of an existing fault and fracture propagation [*Senfaute et al.*, 2003].

In the following we focus on the analysis of the seismicity pattern within a few hours before the collapse. It corresponds to a low tide period for which we can ensure that the recorded signals originated inside the cliff.

The seismic events size is estimated with the signal energy as defined for a digitalized signal [*Evans*, 1979].

$$E=\Sigma A^2 \Delta t, \quad (1)$$

where A is the signal amplitude and $\Delta t$ is the sampling period. The discrete summation is performed for the duration of the transient signal, identifying the beginning and the end of the signal by manual picking. The amplitude A takes into account the sensor sensitivity and the signal amplification in order to have an estimation of the mechanical energy received by the sensor. We observed very good agreement between the energy estimated either by the geophone or by the accelerometer (after signal integration) except for larger events for which the accelerometer signal was saturated. We used the energy for estimating the magnitude by extrapolating of empirical relationship established for induced seismicity. The energy ranges from $10^{-8}$ to $10^{-3}$ J, corresponding to magnitudes ranging from -3.7 to -1.8.

**Seismicity pattern before the Mesnil-Val cliff collapse and data analysis**

Many similarities exist between acoustic emission (AE, usual term at lab scale) and earthquake as revealed by the fact to obey similar statistics over source dimensions spanning more than eight orders of magnitude, possibly ranging from the hundred of kilometers for tectonic earthquakes to the dislocation movement at the atomic size scale [*Miguel et al.*, 2001]. This scaling law, known as the Gutenberg Richter distribution [1954] for earthquakes empirically expresses that the frequency size distribution of brittle fracture is a power law.

$$N(>E) \sim E^{-b} \quad (2)$$

Where $N(>E)$ is the number of events of energy larger than $E$, b is an empirical parameter. For both the accelerometer and the geophone at station 4, we recovered the power law distribution (2) for the event sizes recorded during the two hours preceding the collapse (figure 2). Because we use only one sensor and accordingly did not correct the recorded amplitudes from the event-sensor distance, the strong chalk attenuation may drive a spurious event size distribution. Weiss [1997] have shown that this effect should correspond to (i) a continuous curvature of the size distribution associated to low magnitude order range and (ii) an apparent high b-value, i.e. close to 3. In our case, the observed constant slope over several magnitude orders and the 0.55 exponent value both reject such a possible effect to drive the observed power law distribution of event sizes. To analyze the temporal variation of the b exponent (equation 1), we use moving windows of 100 events with 1 event shift between successive windows (figure 2). We analyzed the event size distribution corresponding to each window (figure 2 presents 1/10 windows for a better readability) and observe that for all the windows the power law distribution is well verified on at least 2 magnitude orders and on more than 4 magnitude orders for the last windows. The b exponent continuously decreases toward the collapse (figure 3).

On the A4 station, the seismic event rate continuously increases until the collapse occurs, reaching the maximal rate of 3 events/s immediately before the collapse. This value corresponds to the maximum event recording rate of the system, i.e. to a continuous recording. This acceleration follows a power law toward time to failure (figure 4) as

$$dN(t)/dt \sim (tc-t)^{-\alpha} \quad (3).$$

dN/dt being the event rate, tc being the failure time, $\alpha$ being a constant.
A similar power law is recovered for the energy acceleration,

$$dE(t)/dt \sim (tc-t)^{-\beta} \quad (4)$$

dE/dt being the energy rate, tc the failure time, $\beta$ a constant.
The exponent value appears to be dependent of the frequency bandwidth of the sensor (figure 4).

**Discussion and conclusion**

The power law increase of the seismic activity before the collapse or the decrease of the b-value (a few studies recovered both patterns simultaneously) is in qualitative agreement with experimental data at the laboratory scale before brittle failure [*Mogi*, 1962; *Scholz*, 1968; *Otani et al.*, 1991; *Lockner*, 1993; *Guarino et al.*, 1998; *Johansen and Sornette*, 2000]. Recently these pattern were reproduced by brittle rupture models and rationalized in the framework of thermodynamics phase transition [*Buchel and Sethna*, 1997; *Zapperi et al.*, 1997; *Sornette and Andersen*, 1998; *Kun and Herrmann*, 1999]. These studies suggest that the macroscopic brittle failure of heterogeneous medium behaves as a critical point in the sense of phase transition. Many observables of the system dynamics are supposed to diverge (as measured by power law behavior) when the system moves toward the critical point [*Sornette*, 2000]. The same framework was recently tested to apply before some earthquakes [*Jaume and Sykes*, 1999]. The power-law behaviour of the seismic activity versus time of our in-situ observations are in agreement with models and with laboratory experiments of two different types: controlled increasing load [*Guarino et al.*, 1998] and controlled constant load, i.e. creep tests, [*Nechad et al.*, 2005]. It argues for the rockfalls and landslides that are gravity driven systems to possibly be candidate to reproduce the critical point model. However the critical point model predicts a constant b-value over time whereas the roll off for large event increases with time [*Jaume and Sykes*, 1999]. This latter effect can result in an apparent b-value decrease. The small number of events we recorded prevents us to further identify from which processes the b-value decrease before the cliff collapse emerges from. This study allows us to interpret the power law precursory pattern before rockfalls or landslides in terms of critical phenomena whereas previously suggested by Voight [1988] to emerge from a tertiary creep acceleration [*Saito and Uezawa*, 1961; *Voight*, 1988]. In an applicative view point, these results can be used for the risk forecasting. The power law acceleration can be recovered in the temporal interval $10^4$-$10^2$ s before the collapse and then be used for determining the time of failure. The b-value decrease is also a possible precursory pattern. In the case of rock slopes, there is few cases [*McCauley*, 1976] for which a seismic monitoring system operated in such a manner that it allowed to identify the precursory behaviors before a collapse. We suggest the present work is the first case identifying a power law acceleration of seismicity rate and energy before a cliff collapse. It opens new routes both to monitor and to understand the physics of rock slope and landslide instabilities. Moreover, it provides new elements for the forecasting of natural structures failure.

**Acknowledgements:** We thank A. Helmstetter, D. Keffer, and F. Lahaie for discussion and useful comments during this study, and the two anonymous reviewers for constructive remarks. This Work was supported by INERIS and the Commission of the European Community (PROTECT Project: EVK3-CT-2000-00029). D. Amitrano was supported by the INSU french grants, Gravitational Instability ACI.

**Figures captions :**

Figure 1 : a) Chak cliff of Mesnil-Val, Normandie coast, France, Microseismic network location and trace of the collapse occurred 23$^{rd}$ june 2002. b) Seismic activity recorded by all the sensors 2 hours before the collapse.

Figure 2: Cumulative distribution of the seismic energy recorded on the geophone (circles) and the accelerometer (squares). Dotted lines give the power-law fit. The exponent interval is given for a 95% confidence level. Continuous lines correspond to successive moving windows (100 events with 10 event shift). The legend indicates the event range relatively to the last event.

Figure 3: b-value during the period preceding the collapse, for successive sliding windows (100 events with 1 event shift), as a function of time to failure (in reversed scale), tc being the time of collapse. The plotted time correspond to the time of the last event of each window. We performed tests on a synthetic catalog for which the b-value is constant (0.55) and the cut-off increases as a power law with the distance from the collapse point. This results in an apparent b-value decrease comparable to the one observed on the experimental data.

Figure 4: Events rate number in event/s (diamonds), and reduced seismic energy rate (dE/dt divided by its maximal value) for the geophone (circles) and for the accelerometer (squares), as function of the time to collapse (tc-t), tc being the time of collapse. The time axis is reversed. The exponent interval is given for a 95% confidence level. To compared our $\beta$ exponent with the Benioff strain acceleration before earthquakes, $\Sigma E^{0.5} \sim (tc-t)^m$, [*Jaume and Sykes*, 1999], m=($\beta$−1)/2= 0.15-0.35.

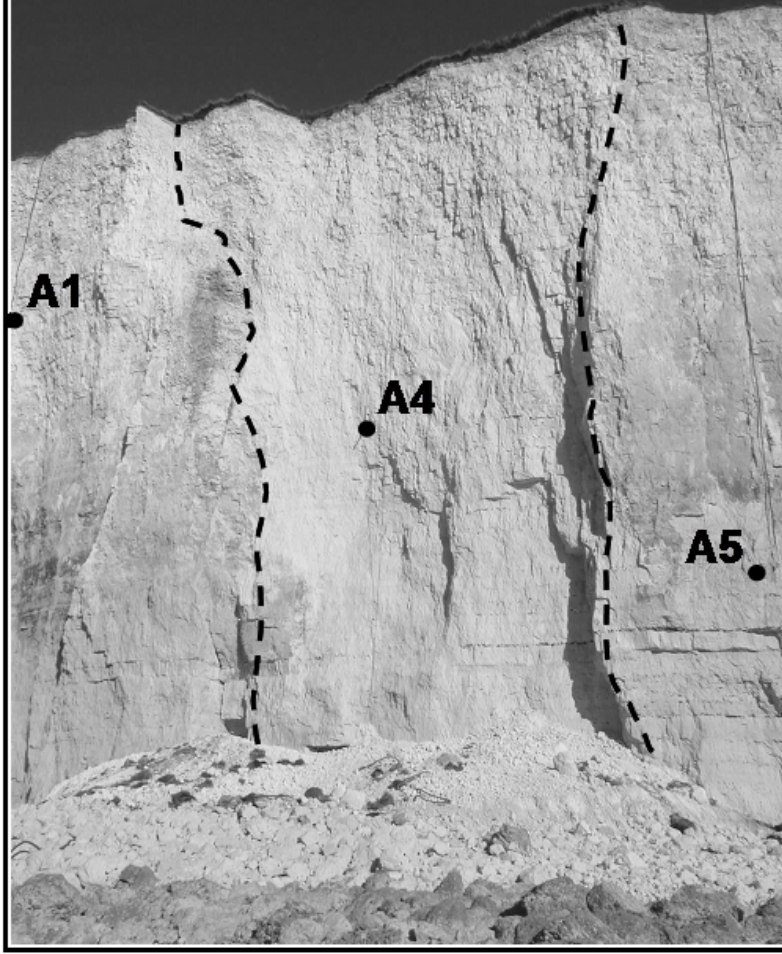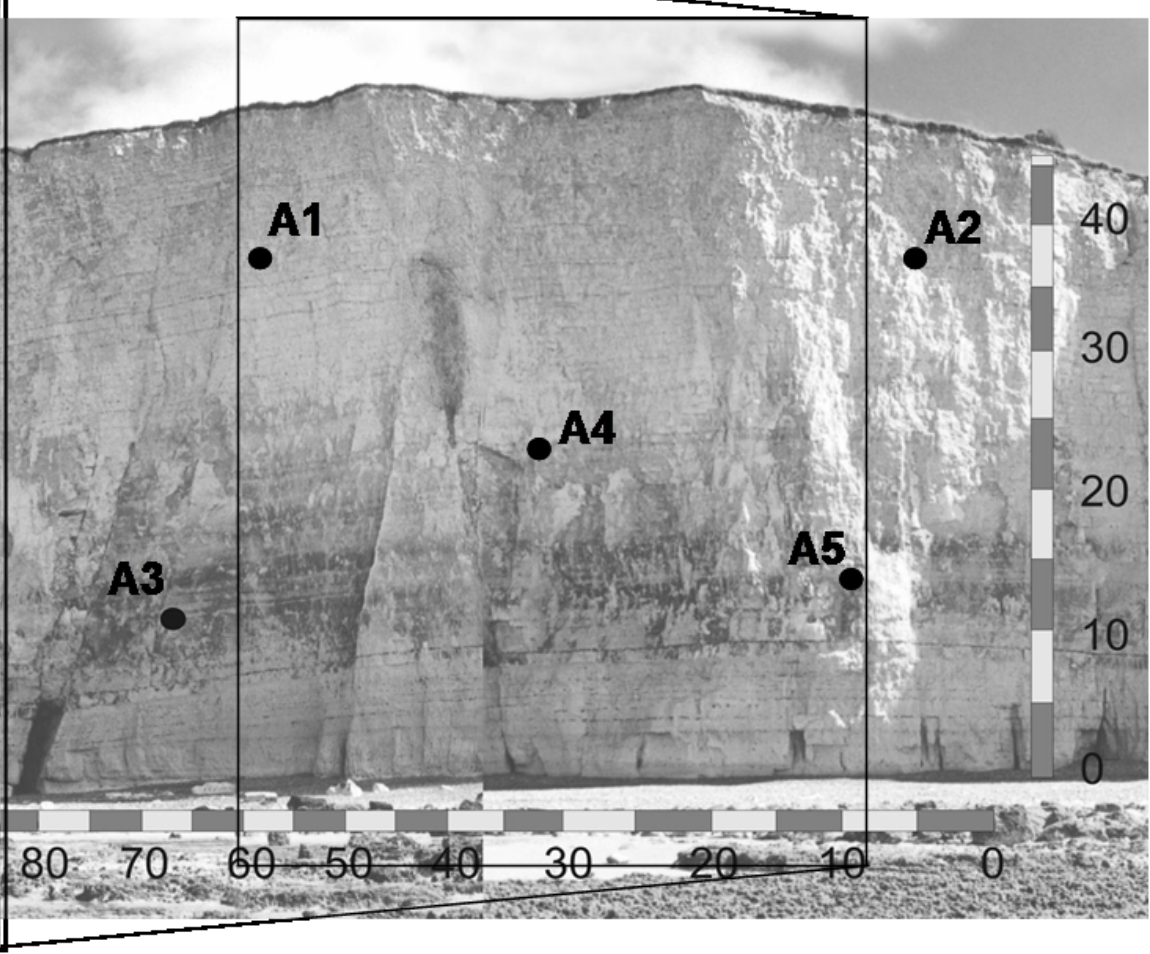

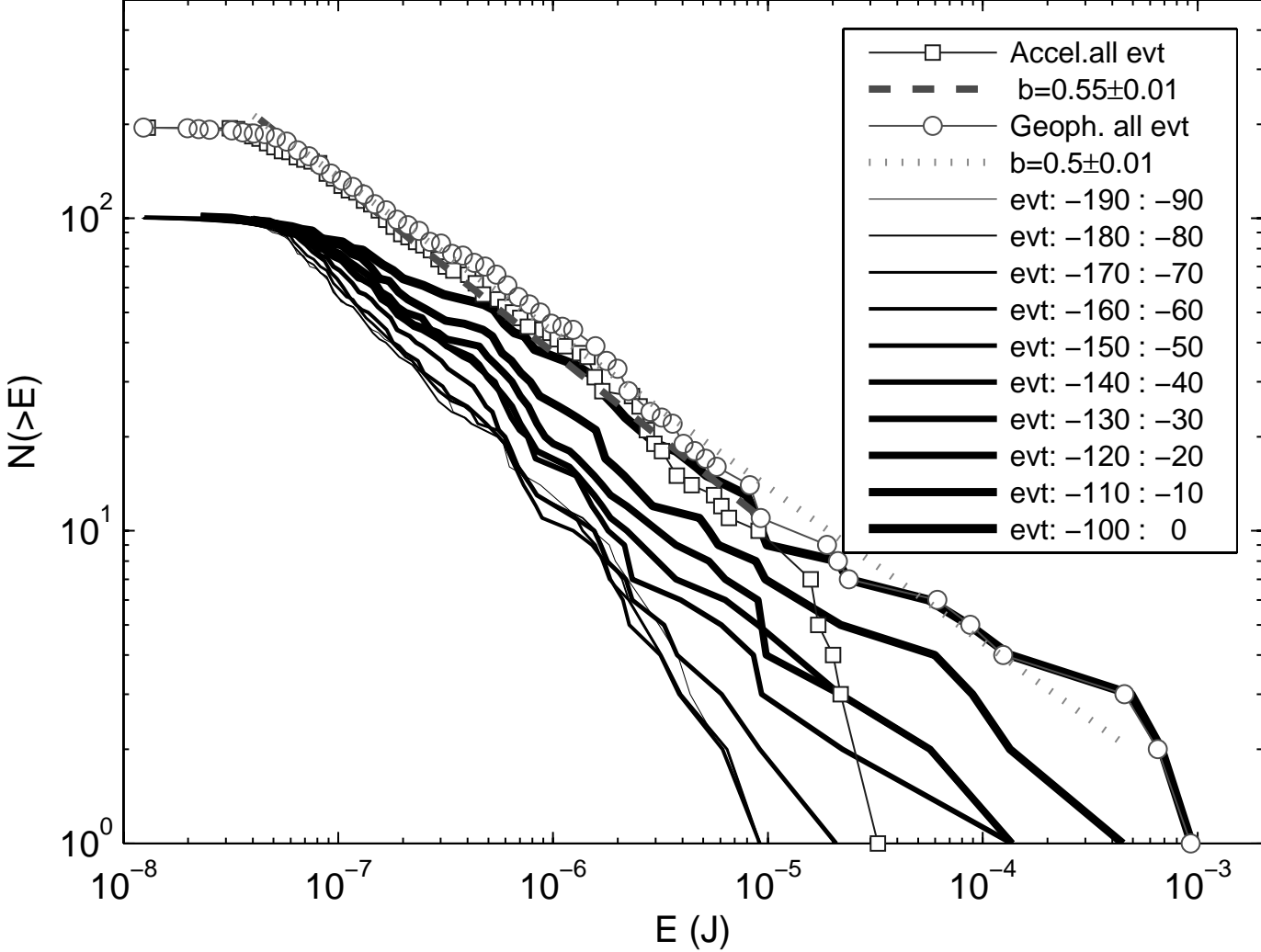

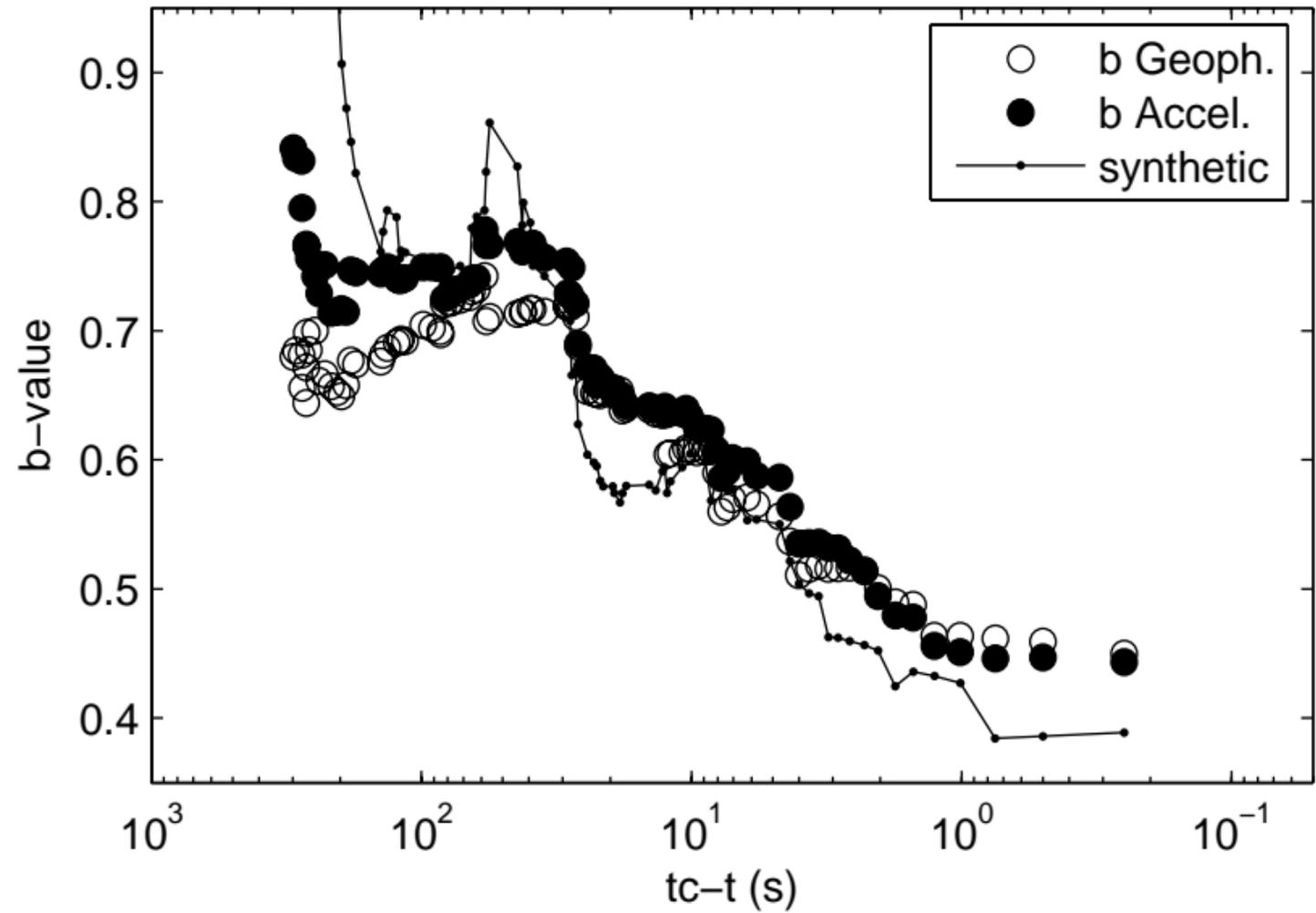

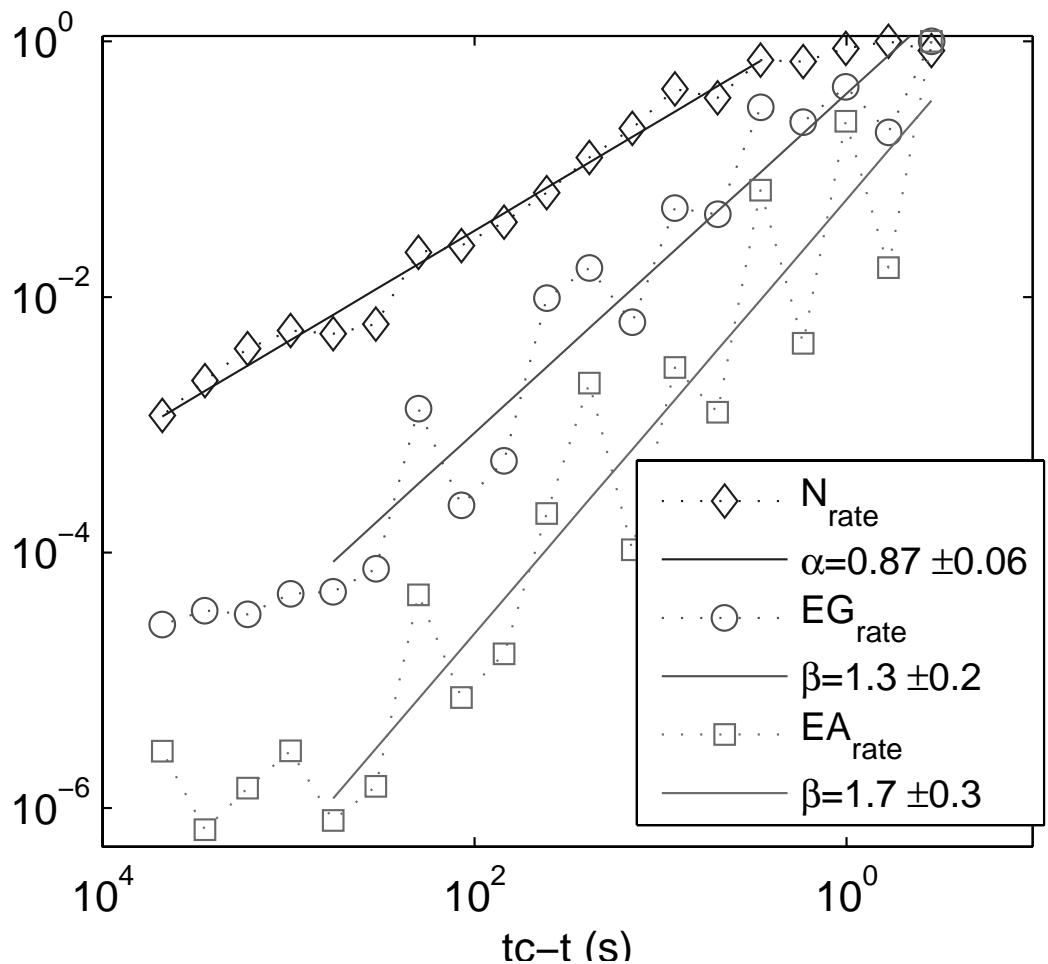